\documentclass[epj]{svjour}
\usepackage{graphics}
\usepackage{amssymb}
\usepackage{color}
\begin{document}
\title{Analogy of QCD hadronization and Hawking-Unruh radiation at NICA}
\author{Abdel Nasser TAWFIK\inst{1,2,3 }
\thanks{http://atawfik.net/}%
}                     
%
%
\institute{
Egyptian Center for Theoretical Physics (ECTP), Modern University for Technology and Information (MTI), 11571 Cairo, Egypt \and 
World Laboratory for Cosmology And Particle Physics (WLCAPP), 11577 Cairo, Egypt \and 
Academy for Scientific Research and Technology (ASRT), Network for Nuclear Sciences (NNS), 11511 Cairo, Egypt
}
\date{Received: 15 October 2015 / Revised version: 17 January 2016}
%
\abstract{
The proposed analogy of particle production from high-energy collisions and Hawking-Unruh radiation from black holes is extended to finite density (collisions) and finite-electric charge (black holes). Assuming that the electric charge is directly proportional to the density (or the chemical potential), it becomes clear that for at least two freezeout conditions; constant $s/T^3$ and $E/N$, the proposed analogy works very well. Dependence of radiation (freezeout) temperature on finite electric-charge leads to an excellent estimation for Kaon-to-pion ratio, for instance, especially in the energy range covered by NICA. The precise and complete measurements for various light-flavored particle yields and ratios are essential in characterizing Hawing-Unruh radiation from charged black holes and  the QCD hadronization at finite density, as well.
} 
\maketitle
\section{Introduction}

A proposal that the black hole gravitational confinement would have an analogy in the hadron color-confinement, dates back to about four decades \cite{Recami:1976}. Accordingly, the QCD vacuum forms an event horizon for the quarks and gluons preventing even quantum tunnelling \cite{0711.3712}. On the other hand, no information is transmitted outwards through the black hole radiation. The Hawking-Unruh radiation \cite{Hawking:1974,Hawking:1975,Unruh:1976} is likely thermal and maintain color neutrality, as well. The universal feature of $e^+\,e^-$, $p-\bar{p}$,  $p-p$ and  $A-A$ collisions that the particle production out of these different scattering processes have thermal patterns  \cite{0711.3712} is puzzling as the $A-A$ thermalization can be interpreted due to the possible rescatterings between the large number of interacting partons, but $e^+\,e^-$ and other elementary collisions not \cite{0711.3712}. The analogy of QCD and black hole does not only solve this thermalization puzzle but proposes a solid interpretation for the phenomenologically proposed  universal freeze-out conditions.
\begin{itemize}
\item First, the vacuum instability through pair production allows quantum tunnelling through the quark and gluon event-horizon. This leads to thermal radiation at a temperature which is determined by the quark-antiquark string tension.
\item Second, such tunnelling processes likely takes place at high energy. Thus, the partition processes lead to a limiting temperature as that in statistical thermal models.
\item Third, the initial state information is likely lost through successive cascades/collisions. This characterizes a kinetic thermalization,  while the stochastic QCD Hawking radiations takes place in equilibrium and this prevents information transfer.
\end{itemize}

The correspondence of the QCD hadronization and Hawking-Unruh radiation can find its phenomenological roots in the gauge theory gravity. Black holes behind which Rindler horizon could confirm the phenomenological (experimental) dependence of temperature ($T$) on the nucleon-nucleon center-of-mass energy ($\sqrt{s_{NN}}$), are Witten black holes, especially for constant $T$, which was observed in all scattering processes at large $\sqrt{s_{NN}}$. An effective description for the string tension ($\sigma$) screening with varying $\sqrt{s_{NN}}$ or baryon chemical potential ($\mu$) has been proposed \cite{exact_string}. The Rindler horizon supports the proposed correspondence between QCD hadronization  and Hawking-Unruh radiation. In the latter, the $q\bar{q}$ pair-production, which is the analogy to the Rindler horizon is a {\it "color-blind"} process caused by color-charge neutrality/confinement. This means that the $q\bar{q}$ pair-production is likely a thermal process because of the random distribution of $q$ and $\bar{q}$, which are entangled in the QCD vacuum.

On the other hand, the produced hadrons are assumed to be {\it "born in equilibrium"}. For hadrons, the Rindler spacetimes are considered as near-horizon approximation of specific black hole spacetimes. Accordingly, the analogy of the thermodynamical features of both QCD hadronization and black hole radiation defines a concrete black hole spacetime  \cite{exact_string}. This was utilized to predict how $T$ depends on $\sqrt{s_{NN}}$ or $\mu$ \cite{exact_string}: (i) the black hole mass should be proportional to $\sqrt{s_{NN}}$ of $\mu$, (ii) the string tension $\sigma$ should be coincident with the effective coupling constant $\alpha_s$ and (iii)  the Hawking temperature should be identical to the Unruh temperature. A fourth condition, namely that the black hole partition function has to diverge at the Hagedorn temperature, was proposed \cite{exact_string}. Accordingly, most of well-known black holes have been disqualified as candidates for the proposed analogy at finite $\sqrt{s_{NN}}$ of $\mu$. We remark that in proposing this last condition, it was assumed that the black hole radiation lasts until the entire evaporation of the black hole mass is completed. A great number of theoretical works examined the conundrum of this Hawking evaporation \cite{giddings}. It was concluded that the evaporation fulfilling causality and thermodynamical consistency should end up with considerable black hole mass remnants. Therefore, the corresponding partition function likely diverges at the Hawking temperature. Such a critical temperature is inversely proportional to the black hole remnant.

\subsection{Charged black-hole radiation and QCD chemical freezeout at finite density}
\label{sec:chrgdBH}

In spacetime of de Sitter-Schwarzschild black hole, the invariant line element can be given as \cite{0704.1426}
\begin{eqnarray}
ds^2 &=& \left(1 - 2\frac{GM}{R_S}\right)~\!dt^2 - \left(1- 2\frac{GM}{R_S}\right)^{-1}\; dr^2,
\label{Schwarz}
\end{eqnarray}
where $r$ and $t$ are flat space- and time-coordinates, respectively.  It is assumed that $G=1/2\sigma=2.63$ and thus  $\sigma=0.19~$GeV$^2$ is the string tension, which can be related to the QCD scale, $\Lambda_{QCD}$ \cite{Tawfik:2015fda}. 

On the other hand, for Reissner-Nordstr\"om black hole with a net electric-charge $Q$, the associated Coulomb repulsion weakens the gravitational attraction, the corresponding event horizon is modified and thus the line element is given as
\begin{eqnarray}
ds^2 &=& \left(1 - 2\frac{G M}{R_{RN}} + \frac{G Q^2}{R_{RN}^2}\right)\; d t^2 \nonumber \\
&-& \left(1- 2\frac{G M}{R_{RN}} + \frac{G Q^2}{R_{RN}^2}\right)^{-1}\; dr^2. 
\label{RN}
\end{eqnarray}
Both interaction field strength and the resulting Reissner-Nordstr\"om radius read
\begin{eqnarray}
f(R_{RN}) &=& 1 - 2 \frac{G M}{R_{RN}} +  \frac{G Q^2}{R_{RN}^2}, \\
R_{RN} &=& \frac{R_S}{2} \left(1 + \sqrt{1 - \frac{Q^2}{G\, M^2}}\right), \label{RNradius}
\end{eqnarray}
where $R_s$ is de Sitter-Schwarzschild black hole radius. It is obvious that $R_{RN}$ is directly proportional to $Q^2$. 

The {\it critical} temperature associating the charged black-hole radiation can be determined from $f'(R_{RN})/4\pi$, where $f'(R_{RN})=\partial f(R_{RN}/\partial R_{RN})$ \cite{book_web},
\begin{eqnarray}
T_{BH}(M,Q) &=& T_{BH}(M,0)\; \left[\frac{4~\left(1 - \frac{Q^2}{G\, M^2}\right)^{1/2}}{\left(1 + \sqrt{1- \frac{Q^2}{G\, M^2}}\right)^{2}}\right], \hspace*{4mm} \label{T-Q}
\end{eqnarray}
where $T_{BH}(M,0)$ is the Hawking-Unruh radiation temperature as determined from de Sitter-Schwarzschild (uncharged) black hole, which was found comparable with the QCD freezeout temperature at vanishing baryon chemical potential, $T_{BH}(M,0) \simeq 175\pm15~$ MeV \cite{1409.3104}. 

A modification in the black hole mass is likely resulted from the Hawking-Unruh radiation
\begin{eqnarray}
d M = T\; d S + \mu\; d Q. \label{eq:1stThrm}
\end{eqnarray}
This is noting but the first law of thermodynamics at finite $\mu$ and electric charge $d Q$ \cite{0007195}. From Bekenstein-Hawking area law, the entropy $S$ reads  \cite{Hawking:1971,Bekenstein}
\begin{eqnarray}
S = \pi \frac{R_{RN}^2}{G}.
\label{entropy}
\end{eqnarray}
A good half-century ago, the profound connections between gravity and thermodynamics was recognized \cite{Cocke1965}. The black hole entropy is considered as Noether charge associated with the diffeomorphism invariance of a classical theory of gravity \cite{Wald1993}.  Recently, it was proposed that gravity can be explained as an entropic force \cite{Verlinde2011}.  Accordingly, the source of thermodynamics (pressure and energy density) including chemical potential and that of gravity are equivalent and concretely gravity could be related to entropy but at finite chemical potential and varying temperature \cite{Young2014}. 

From Eqs. (\ref{RNradius}),  (\ref{eq:1stThrm}) and (\ref{entropy}), the chemical potential was related to the electrically-charged black holes \cite{Tawfik:2015fda,0007195}, 
\begin{eqnarray}
\mu &\simeq & 2\, \pi\, \frac{Q\, R_{RN}}{G\, S} = 2\frac{Q}{R_{RN}}. \label{muQ}
\end{eqnarray}

\subsubsection{Freezeout conditions at high density}

Recently, it has been shown that the freezeout condition, the entropy density normalized to $T^3$, calculated from Hawking-Unruh radiation has almost the same value \cite{1409.3104} as the one proposed for the QCD particle production \cite{Tawfik:2014eba,Tawfik:2005qn,Tawfik:2004ss},
\begin{eqnarray}
\left.\frac{s}{T^3}\right|_{Q=0} = \frac{3}{8\, G^2\, M\, T^3_{BH}(M,0)}. \label{eq:sT3bh}
\end{eqnarray}
Assuming that the proposed analogy of QCD hadronization and Hawking-Unruh radiation is valid at finite density, we show that the constant $s/T^3$ as assumed in QCD hadronization at finite baryon chemical potential \cite{Tawfik:2014eba,Tawfik:2005qn,Tawfik:2004ss} can also be obtained from Hawking-Unruh radiation from the electrically-charged black holes 
 \begin{eqnarray}
\left.\frac{s}{T^3}\right|_{Q \neq 0}=\left.\frac{s}{T^3}\right|_{Q=0} 2 \left(1+\sqrt{1-\frac{Q^2}{G\, M^2}}\right)^5 \left(1-\frac{Q^2}{G\, M^2}\right)^{-3/2}.  \label{eq:sT3b}
 \end{eqnarray}

As also proposed in Ref. \cite{1409.3104}, a second freezeout condition, the average energy per particle, at a vanishing $\mu$ was estimated from de Sitter-Schwarzschild Hawking-Unruh radiation
\begin{eqnarray}
\left.\langle E\rangle/\langle N\rangle\right|_{\mu=0}  &=& \sigma\, R_S. \label{eq:EoverNRs}
\end{eqnarray}
Assuming that the string tension ($\sigma$) in Hawking-Unruh radiation from electrically-charged black holes depends on the chemical potential \cite{exact_string}
\begin{eqnarray}
\sigma(\mu) &\simeq & \sigma(\mu=0)\left[1-\frac{\mu}{\mu_0}\right], \label{eq:sigmu}
\end{eqnarray}
where $\mu_0\simeq 1.2~$GeV.  It is worthwhile to emphasize that at vanishing chemical potential results on the $T$-dependence of the black hole mass in 2-dimensional gravity have been reported and the black-hole mass was related to the center-of-mass energy of QCD collisions \cite{exact_string}. In order to connect the latter with finite baryon-chemical potential, which was proposed from phenomenological studies \cite{Tawfik:2014eba}, a new conserved U($1$) for the baryon charge should be introduced, Eq. (\ref{eq:sigmu}) \cite{exact_string}.  

Eq. (\ref{eq:sigmu}) looks very similar to a lattice QCD parametrization for the chiral temperature at finite baryon chemical potential \cite{FodorJHEP2012}. To account for finite baryon-chemical potential as done in Eq. (\ref{eq:sT3b}) for the universal freezeout condition $s/T^3$, $R_S$ in Eq. (\ref{eq:EoverNRs}) should be replaced by Reissner-Nordstr\"om radius ($R_{RN}$), Eq. (\ref{RNradius})
\begin{eqnarray}
\left.\langle E\rangle/\langle N\rangle\right|_{\mu\neq0} &=& \sigma(\mu)\, R_{RN}.
\end{eqnarray}

\subsubsection{Comparison with thermal models and experimental results} 

\begin{figure}[!htb]
\centering{
\resizebox{0.415\textwidth}{!}{
\includegraphics{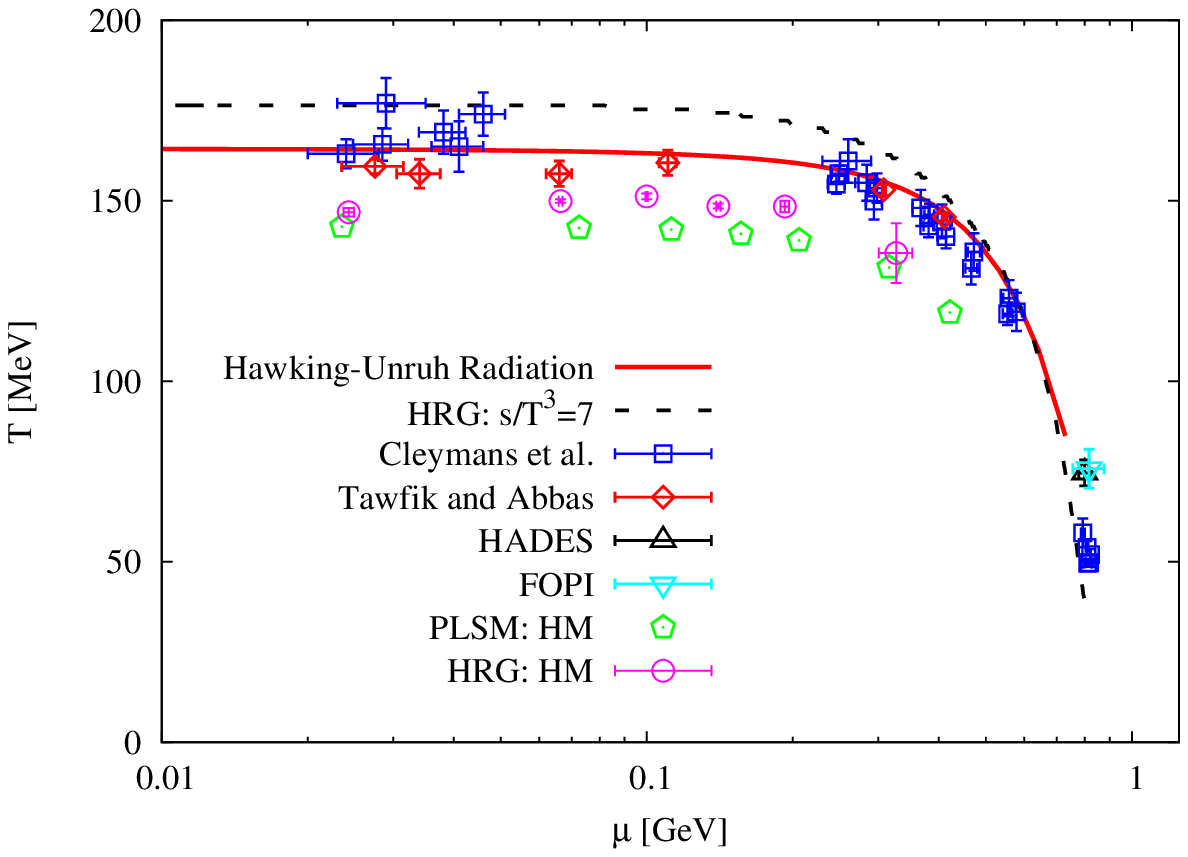} 
}
\resizebox{0.4\textwidth}{!}{
\includegraphics{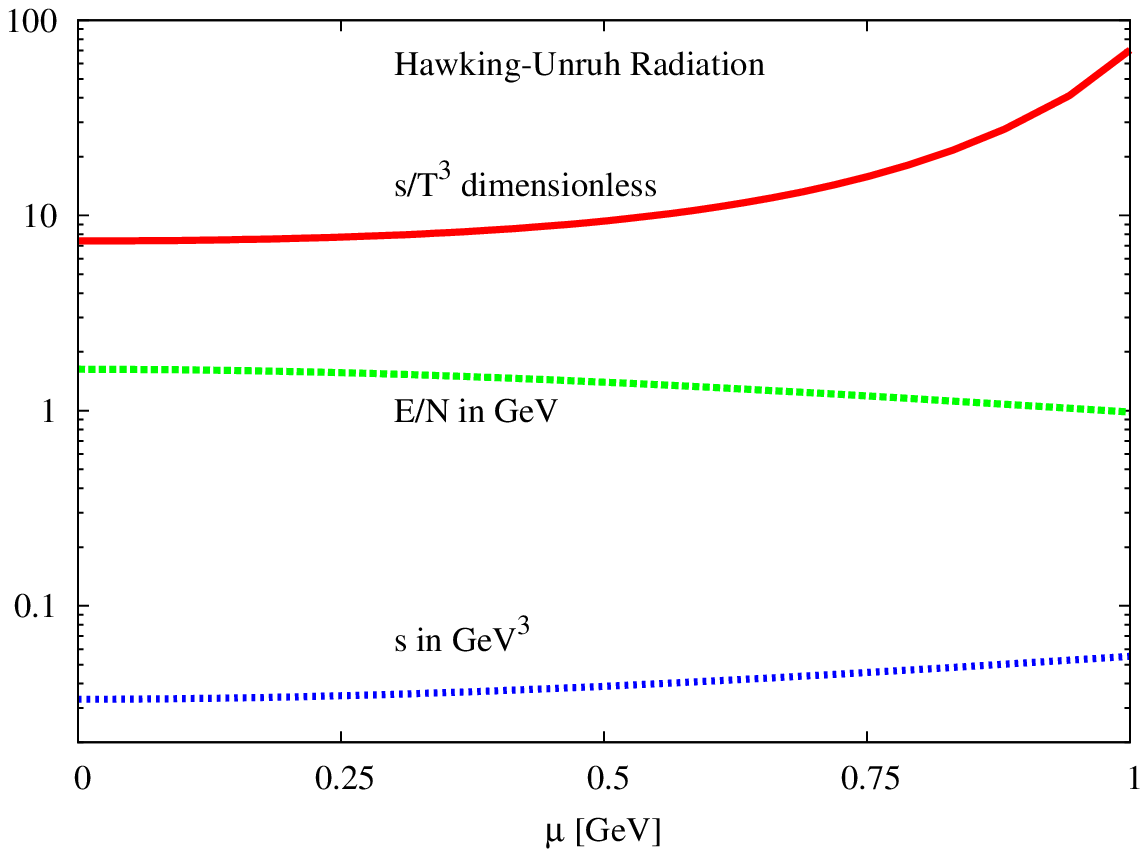}
}\vspace*{0mm}
\caption{Top panel: the analogue of Hawking-Unruh radiation and QCD hadronization is utilized in deducing the freezeout parameters,  $T$, and $\mu_b$ (solid curve). The symbols are the experimentally-determined freezeout parameters from particle ratios: Cleymans {\it et al.} \cite{clmns}, Tawfik and Abbas \cite{Tawfik:2013bza,Tawfik:2014dha}, HADES \cite{hds}, and FOPI \cite{fopi} and higher-order moments:  SU(3) Polyakov linear-$\sigma$ model (PLSM) \cite{AM} and HRG \cite{fohm}. 
Bottom panel: from the proposed analogue, both freezeout conditions $s/T^3$ (solid curve) and $\langle E\rangle/\langle N\rangle$ (dashed curve) and the entropy density $s$ (dotted curve) are given as function of $\mu$.
\label{Fig1} 
}}
\end{figure}

In top panel (a) of Fig. \ref{Fig1}, the freezeout parameters, temperature $T$ and baryon chemical potential $\mu_b$ (solid line), calculated from Eqs. (\ref{T-Q}) and (\ref{muQ}), are compared with the results from HRG model at $s/T^3=7$ (dashed line) and the phenomenological  parameters (symbols) deduced from the particle ratios: Cleymans {\it et al.} \cite{clmns}, Tawfik and Abbas \cite{Tawfik:2013bza,Tawfik:2014dha}, HADES \cite{hds}, and FOPI \cite{fopi} and the higher-order moments of net-proton multiplicity:  SU(3) Polyakov linear-sigma model (PLSM) \cite{AM} and HRG, as well \cite{fohm}. Apart from the slightly lower freezeout temperatures which are determined from the higher-order moments, both qualitative and quantitative comparisons are obviously good. An excellent agreement is obtained with the recent estimation for the freezeout parameters based on RHIC beam energy scan program of the STAR experiment \cite{Tawfik:2013bza,Tawfik:2014dha}. 

It is worthwhile to highlight the excellent agreement at large baryon chemical potential. Almost all experimental results are well reproduced by the black hole thermalization process, which is characterized by Eqs. (\ref{T-Q}) and (\ref{muQ}). The resulting $T$-$\mu$ freezeout phase-diagram is almost identical to the statistical thermal one (dashed curve, for instance) and both describe very well the results from heavy-ion collisions. Here, NICA can play an important role. 

{\bf We recommend precise measurement of various particle yields of light flavors and their ratios at NICA available energies and in all possible interacting systems and centralities. This helps in: 
\begin{itemize}
\item remapping out the $T-\mu$ phase diagram with and without strange quarks,
\item refining the statistical thermal models through confronting them to NICA precise measurements,
\item reexamining the dependence of the black hole radiation temperature on density (charged black holes and black hole remnants),
\item charactering the themalization processes in particle production and black hole radiation at high density, 
\item proposing plausible explanations for other freezeout conditions, and
\item estimating the contributions added by the interactions and the correlations to the thermal model.
\end{itemize}
}

Bottom panel of Fig. \ref{Fig1} shows both freezeout conditions $s/T^3$ (solid curve) and $\langle E\rangle/\langle N\rangle$ (dashed curve) and the entropy density $s$ (dotted curve) as function of  baryon chemical potential. We notice that $s/T^3$ remains constant, $\sim 7$, at $\mu<0.3~$GeV. This is the same range, in which the freezeout temperature remains constant with increasing $\mu$ (top panel) and agrees well with the freezeout temperatures which have been deduced from the thermal fitting of various particle ratios measured in the STAR experiment \cite{Tawfik:2005qn,Tawfik:2004ss}.

From Eq. (\ref{eq:sT3b}),  we realize that $s/T^3$ increases with increasing $\mu$ (or $Q$). When comparing the values of $s/T^3$ from black hole with hadronization, Eq. (\ref{T-Q}), we find that $\left.s/T^3\right|_{\mu \neq 0}$, Eq. (\ref{eq:sT3b}), explains why $\left.T_{BH}\right|_{\mu \neq 0}$ perfectly agrees with the results from the statistical thermal models and from the heavy-ion measurements, while $\left.s/T^3\right|_{\mu \neq 0}\simeq 7$ seems to increase, especially at large chemical potential or black hole charge $Q$.

{\bf NICA precise measurements for particle yields and ratios (i) shed light on the freezeout conditions, (ii) refine the phenomenologically-deduced freezeout parameters ($T$ and $\mu$) and (iii) highlight the role of the phase transition at high density. Furthermore, redrawing the chemical freezeout diagram and charactering the chiral and the deconfinement phase-transitions enables us to distinguish between hadronization, breaking chiral symmetry and the final state of the particle production, especially at high density.}

\subsection{Particle rations and charged black-hole radiation at finite density}

The Unruh temperature is directly proportional to the acceleration, $T=a/2\pi$, which is likely experienced by generated quarks/antiquarks due to the force of the string connecting them to their parents \cite{1502.01938}. At distance $\sim 2 [m_q^2+(\pi \sigma)/2]^{1/2}/\sigma$, the string breaks. For mesons with degenerate quark constituents, the Unruh temperature depends on $m_q$ as follows.
\begin{eqnarray}
T |_{m_q\neq 0} &\simeq & {\sigma \over \pi \sum_q w_q}, \label{UnruhTdegm}
\end{eqnarray}
where the effective mass is given as  \cite{1502.01938}
\begin{eqnarray}
w_q &=& \left[m_q^2 + \frac{\sigma^2}{4 m_q^2 + 2 \pi \sigma}\right]^{1/2}, \label{eq:effectivemass}
\end{eqnarray}
and $m_q$ is the bare mass of $q$-th quark flavor. For massless quarks
\begin{eqnarray}
T |_{m_q=0} \simeq \sqrt{\sigma \over 2\pi}, \label{UnruhTdeg0}
\end{eqnarray}
For nondegenerate quark constituents, the Unruh temperature practically becomes an average of the accelerations of each massive quark. Eqs. (\ref{UnruhTdegm}) - (\ref{UnruhTdeg0}) remain applicable for baryons, as well. At vanishing baryon  chemical potential and different strange quark masses, the corresponding Unruh temperatures for bosons and baryons have been determined \cite{1502.01938}.

In statistical thermal models such as HRG model, the particle number density at finite $T$ and $\mu$ is given as \cite{Tawfik:2014dha,Karsch:2003vd,Karsch:2003zq,Tawfik:2004vv}
\begin{eqnarray}
n(T, \mu) &=& T \sum_{i=1}^{N} \frac{g_i}{2\, \pi^2} \left(\frac{m_i}{T}\right)^2\, \exp\left(\frac{\mu_i}{T}\right)\, K_2\left(\frac{m_i}{T}\right),   \label{eq:n1}
\end{eqnarray}
where $N$ is number of the hadron resonances and $g_i\;$ ($m_i$) being degeneracy (mass) of $i$-th resonance. Straightforwardly, the ratio of two particles, such as $K/\pi$, can be determined. 
\begin{itemize}
\item At vanishing chemical potential
\begin{eqnarray}
\frac{n_K}{n_{\pi}} &=& \left(\frac{m_K}{m_{\pi}}\right)^2 \frac{T_K}{T_{\pi}} \left[\frac{K_2(m_K/T_K)}{K_2(m_{\pi}/T_{\pi})}\right]. \label{eq:nknpi}
\end{eqnarray}
\item At finite baryon chemical potential, both $T_{K}$ and $T_{\pi}$ which can determined from Hawking-Unruh radiation should be extended to finite $\mu$, Eqs. (\ref{T-Q}) and (\ref{UnruhTdegm}). 
\end{itemize}
It is obvious that $K/\pi$ depends on the ratio of their physical masses and of their temperatures. The latter can be deduced from Eqs. (\ref{T-Q}) and (\ref{UnruhTdegm}) and is apparently related to the effective and bare masses of their quark constituents. As introduced in Ref. \cite{1502.01938}, Eq. (\ref{eq:nknpi}) counts for the primary production for kaon and pion particles. This is the case in Hawking-Unruh radiation. In thermal models, the break-up of $\bar{q}q$-strings of the hadron resonances is assumed to occur, arbitrarily, i.e. each string breaks up into two substrings, which might form stable hadron or resonance or a large-mass string \cite{KWerner1990}. This fragmentation process follows the Bootstrap pattern and ends up with an ordinary resonance decay. 
\begin{eqnarray}
\frac{n_K}{n_{\pi}} &=& \frac{\sum_i\left[\left.n_K\right|^{stable}_i + \sum_{j\neq i} b_{j\rightarrow i} (n_K)_j\right]}{\sum_k\left[\left.n_{\pi}\right|^{stable}_k + \sum_{l \neq k} b_{l\rightarrow k} (n_{\pi})_k\right]},
\end{eqnarray}
where $b_{\cdots}$ stand for the branching ratios of the given decay channels.

\begin{figure}[!htb]
\centering{
\resizebox{0.45\textwidth}{!}{
\includegraphics{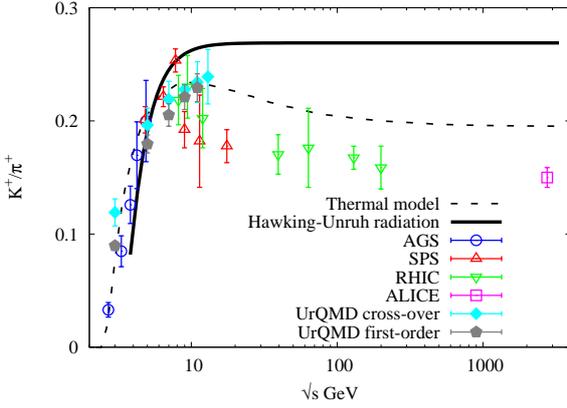}
}\vspace*{0mm}
\caption{$K^+/\pi^+$ as function of $\sqrt{s_{NN}}$ is calculated from statistical thermal model (dashed curve) and Hawking-Unruh radiation (solid curve) and compared with various experimental results (open symbols) and UrQMD assuming cross-over and first-order phase transition (solid symbols). 
\label{BH_Kp_pip}
}}
\end{figure}

The results are depicted in Fig. \ref{BH_Kp_pip} (thick curve). The authors of \cite{1502.01938} estimated $K/\pi$ at a vanishing chemical potential. It is obvious that our calculations confirm this (at extremely large $\sqrt{s_{NN}}$). Fig. \ref{BH_Kp_pip} presents an extension to the proposal introduced in Ref. \cite{1502.01938}. Here, the dependence of $K^+/\pi^+$ on $\sqrt{s_{NN}}$ is calculated from statistical thermal model (dashed curve) and Hawking-Unruh radiation (solid curve) and compared with various experimental results (open symbols) and with UrQMD, in which cross-over and first-order phase-transition are assumed (solid symbols). The excellent agreement at $\sqrt{s_{NN}}\lesssim 10~$GeV highlights the strong correspondence between the statistical thermal models and the Hawking-Unruh thermalization and between both approaches and the experimental results. This range of energy is well accessed by the NICA future facility, which offers a precise energy scan in various energies, interacting systems, centralities, etc. with high luminosity. {\bf In light of this, we highlight that the NICA precise measurement of various particle yields would shed more light to our understanding of the black hole radiation at high density.}

On the other hand, we notice that the particle ratios at large energies (small baryon chemical potentials) are considerably overestimated. Both statistical thermal model and black hole radiation do not fit well with the experimental results. One possible interpretation would be the quark occupation factors, which are entirely omitted in the present calculations, Eq. (\ref{eq:nknpi}). In additional to this, the main assumption of the statistical thermal model, the absence of interactions, would also be responsible even partly for such overestimation.

%

\end{document}